*Article*

# Scrutinizing dark-matter scenarios with $B \to (K, K^*)\bar{\nu}\nu$ decays

Alexander Berezhnoy [1], Wolfgang Lucha [2] and Dmitri Melikhov [3]

[1] D. V. Skobeltsyn Institute of Nuclear Physics, M. V. Lomonosov Moscow State University, 119991, Moscow, Russia; A.V.Berezhnoy@gmail.com

[2] Institute for High Energy Physics, Austrian Academy of Sciences, Nikolsdorfergasse 18, A-1050 Vienna, Austria; Marietta Blau Institute (MBI) for Particle Physics, Dominikanerbastei 16, A-1010 Vienna, Austria; Wolfgang.Lucha@oeaw.ac.at

[3] D. V. Skobeltsyn Institute of Nuclear Physics, M. V. Lomonosov Moscow State University, 119991, Moscow, Russia; Joint Institute for Nuclear Research, 141980 Dubna, Russia; Faculty of Physics, University of Vienna, Boltzmanngasse 5, A-1090 Vienna, Austria; dmitri_melikhov@gmx.de

**Abstract:** Conceivable explanations of Belle-II measurements of a (surprising) *excess* of missing-energy decays of the *B* meson to the *K* meson not covered by standard-model neutrino–antineutrino pairs might be offered by additional contributions of dark-matter fermion–antifermion pairs. Assuming the excessive missing-energy events to be mediated by a (generic) scalar or vector boson, a simultaneous inspection of both the missing-energy *B* decays into a pseudoscalar *K* meson or a vector $K^*$ meson enables to gain information on the nature of bosons relating standard-model and dark-matter sectors, irrespective of any (unknown) dark-sector details. Upon availability of indispensable experimental data, most prominent among such insights might be the identification of the mediator spin from the differential *B*-meson decay widths.

## 1. Impetus: (Hardly Expected) Detection of Missing-Energy Excess Events

In November 2023, already nearly two years ago, the Belle II collaboration announced a *measurement* of the decay of the pseudoscalar bottom meson, $B^+$, into a pseudoscalar strange meson $K^+$ and missing energy $M_X$ (also referred to as $B^+ \to K^+ \bar{\nu}\nu$), with branching ratio [1]

$$\mathcal{B}(B^+ \to K^+ \bar{\nu}\nu) = (2.3 \pm 0.7) \times 10^{-5} = (5.4 \pm 1.5)\, \mathcal{B}(B^+ \to K^+ \bar{\nu}\nu)_{\rm SM} \,. \qquad (1)$$

Astonishingly, the branching ratio (1) is, in fact, several times larger than expected within the standard model (SM) for this kind of missing-energy decay into a neutrino–antineutrino pair and, in addition, in (possibly negligible) conflict with the corresponding upper bound on the branching ratio of the latter decay, established a few years earlier by the Belle experiment [2],

$$\mathcal{B}(B^+ \to K^+ \bar{\nu}\nu) < 1.9 \times 10^{-5} \,.$$

The Belle-II experimental finding (1) immediately triggered intense or hectic theoretical activity [3–31], aiming at explanations of these missing-energy excess events by hypothetical particles of different sorts, among which there also is, rather prominently, dark matter (DM). Instead of dealing with a particular DM scenario, our intention [5,25,32,33] is to demonstrate how to unambiguously rule out entire classes of DM models by (maybe future) experiments.

(Accordingly, we refrain from discussing the many variants of proposed explanations of the observed excess events, as this would merely result in a distraction from the main aspects of the present considerations and, very likely, considerable confusion on the side of the reader. The two specific DM models of genuine interest to us are described, in due detail, in Sect. 3.)

More precisely, our central goal is to show — for the example of two representatives of two particular categories of DM models, separately discussed in Ref. [25] and Ref. [32], respectively, and recalled in Sect. 3 — that the preferability of one of these DM scenarios over the other may be decided — without knowledge of the details of the DM models under scrutiny — by inspecting simultaneously for both options some experimentally measurable quantities, introduced in Sect. 4 and related to the differential or integrated widths of both of the missing-energy $B$-meson decays $B \to K M_X$ and $B \to K^* M_X$. To this end, we compile, within the present investigation, the actual findings of the two previous analyses [25,32] — which, to a large part, are of primarily graphical nature — and combine these outcomes under a unique common notation, leaving aside all of the technical details of the derivations, which have been given in Refs. [25,32]. This move should facilitate both the treatment of the two bulks of information in parallel and their systematic confrontation on an equal footing.

From our point of view, the reasonable realization of this (not too ambitious) program designed above necessitates, however, to reproduce in the present paper a variety of figures of particular relevance from earlier publications [25,32] in precisely the same shape as given in Refs. [25,32]. (As a matter of fact, it is our conviction that every attempt to discuss, for instance, differences between the functional dependencies of predictions for some quantity of interest without having at hand the related plots, and to merely refer to the corresponding figures in one's earlier publications, has to be regarded as inadequate and blatant nonsense.)

## 2. Towards Dark-Matter Resolutions of the Missing-Energy Excess Puzzle

Let us focus to (prototypic) DM models based on any mediator boson $R$ that couples to both SM particles and DM particles and, accordingly, may act as a kind of interface between the SM sector, on the one hand, and the DM sector, on the other hand. The DM particles of interest are assumed to be a generic fermion $\chi$ and its antifermion $\bar\chi$. Consequently, any such mediator boson $R$ might induce missing-energy decays of the $B^+$ meson to the pseudoscalar $K^+$ meson ($B^+ \to K^+ M_X$) or to the vector $K^{*+}$ meson ($B^+ \to K^{*+} M_X$), where a pair of DM fermion $\chi$ and antifermion $\bar\chi$ is responsible for part of the missing energy. These decays proceed (dominantly) via the flavour-changing neutral-current process $b \xrightarrow{R} s\bar\chi\chi$ sketched in Fig. 1: Enabled by one internal exchange of a charged SM gauge boson $W^-$, the mediator $R$ couples to the top quark $t$ as well as to the DM fermion–antifermion pair $\bar\chi\chi$ in the final state.

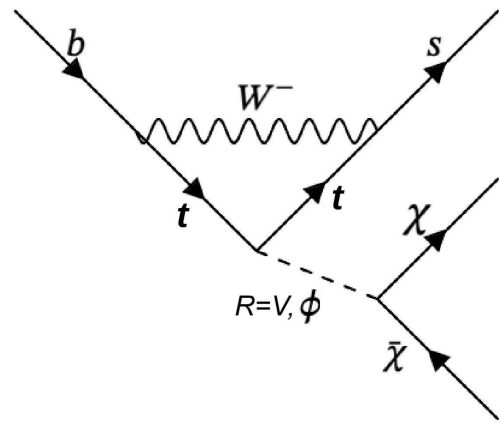

**Figure 1.** Penguin-type $W$-boson–$t$-quark loop diagram [32], effecting the $B$-meson decay $B \xrightarrow{R} K^{(*)} \bar\chi\chi$ into a $K^{(*)}$ meson plus a dark fermion–antifermion pair $\bar\chi\chi$ by means of intermediate $\phi$ or $V$ mediators.

In this context, particular importance has to be conceded to just those particle types that are able to act as mediators between the (fermionic) SM sector and the (fermionic) DM sector, viz., to all scalar or vector bosons that (might) couple to both SM fermions and DM fermions. In the following, the two kinds of mediator boson will be referred to by $\phi$ if of scalar type and by $V$ if of vector type; both will collectively be subsumed under the label $R$, that is, $R = \phi, V$.

The missing energy $M_X$ encountered in decays of the kind $B^+ \to K^{(*)+} M_X$ is shared by all decay products that escape experimental detection and thus determined by the difference

$$q \equiv p_B - p_{K^{(*)}}$$

of the $B$-meson momentum $p_B$ and the $K^{(*)}$-meson momentum $p_{K^{(*)}}$ *observed* by experiment:

$$M_X = \sqrt{q^2} \equiv \sqrt{(p_B - p_{K^{(*)}})^2} \ .$$

Within the present approach, for both $B$-meson missing-energy decays $B^+ \to K^{(*)+} M_X$ two decay channels are available: to either a pair of SM neutrino $\nu$ and antineutrino $\bar\nu$ or a pair of DM fermion $\chi$ and antifermion $\bar\chi$. For the missing energy $M_X$, this implies the two options

$$q \equiv p_B - p_{K^{(*)}} = p_\nu + p_{\bar\nu} \quad \Longrightarrow \quad M_X^2 = q^2 \equiv (p_B - p_{K^{(*)}})^2 = (p_\nu + p_{\bar\nu})^2 \ ,$$
$$q \equiv p_B - p_{K^{(*)}} = p_\chi + p_{\bar\chi} \quad \Longrightarrow \quad M_X^2 = q^2 \equiv (p_B - p_{K^{(*)}})^2 = (p_\chi + p_{\bar\chi})^2 \ .$$

The partial decay widths for $B \to K^{(*)} \bar\nu \nu$ and $B \xrightarrow{R} K^{(*)} \bar\chi \chi$ combine to the total decay widths

$$\Gamma(B \to K^{(*)} M_X) = \Gamma(B \to K^{(*)} \bar\nu \nu)_\text{SM} + \Gamma(B \xrightarrow{R} K^{(*)} \bar\chi \chi) \ , \qquad R = \phi, V \ .$$

## 3. Representatives of DM Proposals Using Intermediate Mediator Bosons

Now, of particular interest to us must be the Lagrangian $\mathcal{L}_R$ describing the couplings of a mediator $R$ to the top quark $t$ and to DM fermions $\chi$ and, likewise, the *effective* Lagrangian $\mathcal{L}_{b \to sR}$ encoding, in form of an effective coupling parameter $g_{bsR}$, the effective interaction of the mediator boson $R$ with the $b$ quark of the $B$ meson and the $s$ quark of the $K^{(*)}$ meson.

For mediator-boson masses $M_R$ larger than twice the mass $m_\chi$ of the dark fermion $\chi$, $M_R > 2\,m_\chi$, all mediators $R = \phi, V$ unavoidably develop nonvanishing, and $q^2$-dependent, decay widths of the decay into a DM fermion $\chi$ and antifermion $\bar\chi$, $R \to \bar\chi \chi$. Upon assuming such a partial decay width to form the dominant contribution to the *total* decay width $\Gamma_R(q^2)$ of the mediator boson $R$, the latter decay width may be straightforwardly extracted from the full propagator of the mediator $R$. It proves to be a product of a $q^2$-independent factor $\Gamma_R^0$ of second order in the coupling parameter of the mediator $R$ to a DM fermion–antifermion pair, $g_{R\chi\chi}$, and a $q^2$-dependent factor $\Delta_R(q^2)$ which is nonzero just for $q^2$ values greater than $4m_\chi^2$:

$$\Gamma_R(q^2) = \Gamma_R^0 \, \Delta_R(q^2) \ , \qquad R = \phi, V \ . \tag{2}$$

Each factor involves both the mass $M_R$ of the mediator $R$ and the mass $m_\chi$ of the fermion $\chi$.

Among others, our analysis [5,25,32,33] relies on the simultaneous inspection of the DM-involving $B$ decays $B \xrightarrow{R} K \bar\chi \chi$ and $B \xrightarrow{R} K^* \bar\chi \chi$ and exploits the differential decay widths

$$\frac{d\Gamma(B \xrightarrow{R} K^{(*)} \bar\chi \chi)}{dq^2} \ , \qquad K^{(*)} = K, K^* \ , \qquad R = \phi, V \ . \tag{3}$$

The explicit expressions of the two differential distributions for the two (exemplary) DM models recalled in Subsects. 3.1 and 3.2 are indeed a little bit lengthy; they may be found, for the scalar mediators $R = \phi$, in Refs. [25,32] and, for the vector mediators $R = V$, in Ref. [32].

*3.1. Example of a Top-philic Dark-Matter Scenario Incorporating Mediator Bosons of Scalar Nature*

A rather popular DM model [34,35] with a *scalar* mediator $\phi$ is defined by its interaction Lagrangian encompassing the coupling of $\phi$ to the *t* quark — usually formulated in terms of the Higgs-boson vacuum expectation value $v$ ($\cong$ 246 GeV), the *t* mass $m_t$, and a parameter $y$ — and to the DM fermion $\chi$ as well as an *effective* Lagrangian which captures the *b*-*s*-$\phi$ vertex:

$$\mathcal{L}_\phi = -\frac{y\,m_t}{v}\,\bar{t}t\,\phi - g_{\phi\chi\chi}\,\bar{\chi}\chi\,\phi\,, \tag{4}$$

$$\mathcal{L}_{b\to s\phi} = g_{bs\phi}\,\bar{s}_L b_R\,\phi\,. \tag{5}$$

Within this model, the factors combining to the decay width (2) of the mediator $\phi$ are [36,37]

$$\Gamma_\phi^0 = \frac{g_{\phi\chi\chi}^2}{8\pi}\,M_\phi\left(1-\frac{4m_\chi^2}{M_\phi^2}\right)^{\frac{3}{2}}, \tag{6}$$

$$\Delta_\phi(q^2) = \left(\frac{q^2-4m_\chi^2}{M_\phi^2-4m_\chi^2}\right)^{\frac{3}{2}}\frac{M_\phi}{\sqrt{q^2}}\,\Theta(q^2-4m_\chi^2)\,.$$

The hadronic $B \overset{R}{\leftrightarrow} K^{(*)}$ transition amplitudes entering the *B*-meson decay widths may be parametrized in terms of two (dimensionless) form factors $f_0^{B\to K}(q^2)$ and $A_0^{B\to K^*}(q^2)$ [38]:

$$\langle K|\bar{s}(1-\gamma_5)b|B\rangle = \langle K|\bar{s}b|B\rangle = \frac{M_B^2-M_K^2}{m_b-m_s}\,f_0^{B\to K}(q^2)\,,$$

$$\langle K^*|\bar{s}(1-\gamma_5)b|B\rangle = -\langle K^*|\bar{s}\gamma_5 b|B\rangle = -i\,(\varepsilon q)\,\frac{2\,M_{K^*}}{m_b+m_s}\,A_0^{B\to K^*}(q^2)\,.$$

Convenient parametrizations [39] of these hadronic form factors, obtained by grasping the nonperturbative QCD effects by lattice QCD [40] and light-cone sum rules [41,42], are given (together with their counterparts required by models with vector mediators) in Appendix A.

*3.2. Example of a Top-philic Dark-Matter Scenario Incorporating Mediator Bosons of Vector Nature*

In a very similar manner, a DM model [43–45] that features some *vector* mediator $V$ — which couples to both vector and axialvector quark and dark-fermion currents — exploits an interaction Lagrangian consisting of the coupling of $V$ to the *t* quark (with parameter $g_{Vtt}$) and to the DM fermion $\chi$, and a (resulting) *effective* Lagrangian representing the *b*-*s*-*V* vertex:

$$\mathcal{L}_V = g_{Vtt}\,\bar{t}\gamma_\mu(1+\gamma_5)t\,V^\mu + g_{V\chi\chi}\,\bar{\chi}\gamma_\mu\chi\,V^\mu\,, \tag{7}$$

$$\mathcal{L}_{b\to sV} = g_{bsV}\,\bar{s}\gamma_\mu(1-\gamma_5)b\,V^\mu\,. \tag{8}$$

The two factors the product of which forms the decay width (2) of the mediator $V$ are [33,46]

$$\Gamma_V^0 = \frac{g_{V\chi\chi}^2}{12\pi}\,M_V\sqrt{1-\frac{4m_\chi^2}{M_V^2}}\left(1+\frac{2m_\chi^2}{M_V^2}\right), \tag{9}$$

$$\Delta_V(q^2) = \frac{M_V}{\sqrt{q^2}}\sqrt{\frac{q^2-4m_\chi^2}{M_V^2-4m_\chi^2}}\,\frac{q^2+2m_\chi^2}{M_V^2+2m_\chi^2}\,\Theta(q^2-4m_\chi^2)\,.$$

## 4. Telling Scalar-Mediator DM Models from Vector-Mediator DM Models

Before embarking on the intended revision of somewhat scattered earlier findings, we feel strongly obliged to repeat, at this place, our statement in the Introduction that — for all the reasons explained in great detail in the second half of the Introduction — we regard as unavoidable to reproduce, without changes, a couple of figures that, with the exception of Fig. 6, have already been presented in previous publications. (It goes without saying that the origin of these figures is unambiguously identified in the respective captions of the figures.)

*4.1. Ratios of (Differential) B-Meson Decay Widths as Basic Tools for a Discrimination of Mediators*

For our investigation in parallel, not confined to any specific DM model, of the *B* decays

$$B \xrightarrow{R} K^{(*)} \bar\chi\chi \,, \qquad K^{(*)} = K^+, K^{*+} \,, \qquad R = \phi, V \,, \tag{10}$$

two conveniently defined expressions [5,25,32,33] prove to be of great importance. These are

1. the ratio of the *differential* decay widths (3) of both missing-energy *B*-meson decays (10)

$$\widehat{\mathcal{R}}^{(R)}_{K^*/K}(q^2) \equiv \frac{\dfrac{d\Gamma(B \xrightarrow{R} K^* \bar\chi\chi)}{dq^2}}{\dfrac{d\Gamma(B \xrightarrow{R} K \bar\chi\chi)}{dq^2}} \,, \qquad R = \phi, V \,; \tag{11}$$

2. the ratio of the integrated decay widths, $\Gamma$, of both missing-energy *B*-meson decays (10)

$$\mathcal{R}^{(R)}_{K^*/K} \equiv \frac{\Gamma(B \xrightarrow{R} K^* \bar\chi\chi)}{\Gamma(B \xrightarrow{R} K \bar\chi\chi)} \,, \qquad R = \phi, V \,. \tag{12}$$

Their scrutiny allows for possibly advantageous insights into the DM scenarios under study:

- For a given mediator boson $R = \phi, V$, the DM parameters enter in the differential decay widths (3) by a factor common to both *B*-meson decays (10), which necessarily cancels exactly in the respective differential decay-width ratio (11). Consequently, the ratio (11) does not depend on (and is totally insensitive to) the numeric values of DM parameters.
- The differential decay-width ratio (11) turns out to be highly sensitive to the spin of the boson *R* actually mediating the missing-energy decays (10), as evident from Fig. 2: For scalar mediators $R = \phi$, this ratio drops, slowly but monotonously, from $\widehat{\mathcal{R}}^{(\phi)}_{K^*/K}(0) \gtrapprox 1$ to its zero at $M_X/(M_B - M_{K^*}) = 1$. For vector mediators $R = V$, this ratio first grows from $\widehat{\mathcal{R}}^{(V)}_{K^*/K}(0) = 1$ to its maximum (near $M_X/(M_B - m_{K^*}) \approx 0.93$) and then decreases rather quickly to its zero at $M_X/(M_B - M_{K^*}) = 1$. This difference in behaviour offers a tool to identify, via the spin of the involved mediator, viable categories of DM models.

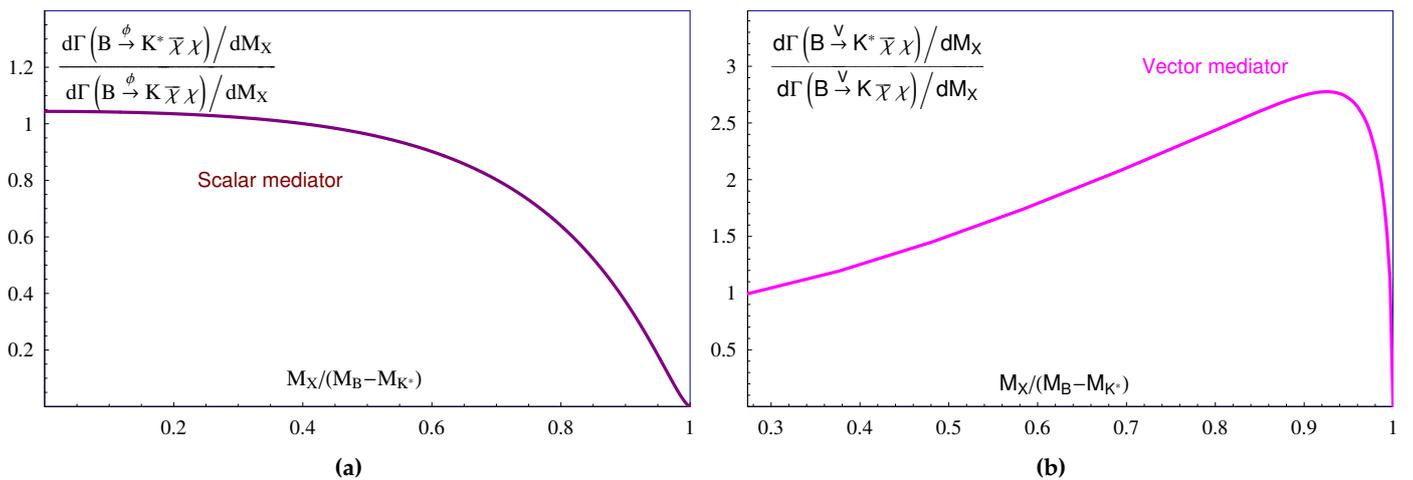

**Figure 2.** *Differential-width* ratio (11) vs. missing mass $M_X$, for (**a**) scalar [25] or (**b**) vector [32] mediator.

- The integrated decay-width ratio (12) exhibits a rather pronounced dependence on the mass $M_R$ — and, to a lesser extent, on the decay width $\Gamma^0_R$ — of the mediator $R = \phi, V$, as illustrated by Fig. 3. More precisely, for all mediator masses $M_\phi$ or $M_V$, respectively, this ratio is less than 1 for a scalar mediator $\phi$ and larger than 1 for a vector mediator $V$:

$$\mathcal{R}^{(R)}_{K^*/K}(M_R)\begin{cases} \lesssim 1 & \forall\, M_\phi & \text{for scalar mediators } R = \phi\,, \\ > 1 & \forall\, M_V & \text{for vector mediators } R = V\,. \end{cases}$$

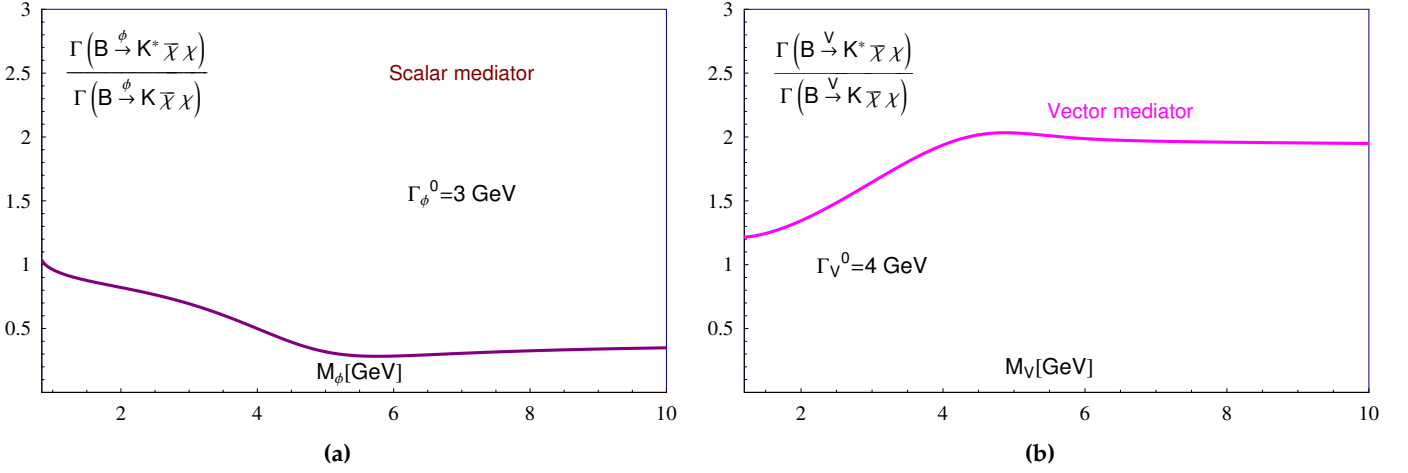

**Figure 3.** *Integrated-width* ratio (12) vs. mediator mass $M_{\phi,V}$ [32] for (**a**) scalar and (**b**) vector mediators.

*4.2. Further Contemplations Taking Advantage of Suitable Ratios of B-Meson Partial Decay Widths*

The existing constraints on acceptable DM scenarios may be even severely tightened by confronting the width of the missing-energy decay $B \to K^* M_X$ with that of the (SM-driven) decay $B \to K^* \bar\nu\nu$ by considering, in particular, the ratio of these two partial decay widths,

$$\widetilde{\mathcal{R}}^{(R)}_{K^*} \equiv \frac{\Gamma(B \to K^* M_X)}{\Gamma(B \to K^* \bar\nu\nu)_{\text{SM}}}\,, \tag{13}$$

and reformulating it as a product of ratios of $B$-decay widths that have been either predicted by theory or measured by experiment and thus may be regarded as (rather) well established:

$$\widetilde{\mathcal{R}}^{(R)}_{K^*} = 1 + \mathcal{R}^{(R)}_{K^*/K}\,\frac{\Gamma(B \to K\bar\chi\chi)}{\Gamma(B \to K\bar\nu\nu)_{\text{SM}}}\,\frac{\Gamma(B \to K\bar\nu\nu)_{\text{SM}}}{\Gamma(B \to K^*\bar\nu\nu)_{\text{SM}}}\,.$$

For the third factor, the SM branching ratios [47–49] of the $B$ decays to $K^{(*)}$ and a $\bar\nu\nu$ pair,

$$\begin{aligned}\mathcal{B}(B^+ \to K^+ \bar\nu\nu) &= (4.44 \pm 0.30) \times 10^{-6}\,, \\ \mathcal{B}(B^+ \to K^{*+} \bar\nu\nu) &= (9.8 \pm 1.4) \times 10^{-6}\,,\end{aligned} \tag{14}$$

and, for the second factor, our $\bar\chi\chi$ "interpretation" inferred from the Belle-II [1] result (1),

$$\Gamma(B^+ \to K^+ \bar\chi\chi) = (4.4 \pm 1.5)\,\Gamma(B^+ \to K^+ \bar\nu\nu)_{\text{SM}}\,,$$

may be adopted, in order to end up with some relationship between the ratios (13) and (12):

$$\widetilde{\mathcal{R}}^{(R)}_{K^*} = 1 + (2 \pm 0.6)\,\mathcal{R}^{(R)}_{K^*/K}\,. \tag{15}$$

Moreover, combining the Belle upper limit on the branching ratio of the decay $B \to K^* \bar\nu\nu$ [2]

$$\mathcal{B}(B \to K^* \bar\nu\nu) < 2.7 \times 10^{-5}$$

with the branching ratio (14), and its uncertainty, provides an upper bound on the ratio (13):

$$\widetilde{\mathcal{R}}_{K^*}^{(R)} < 3.2 \ . \tag{16}$$

Needless to stress, the relation (15) and the upper bound (16) must be compatible. This reasonable demand may impose additional constraints on both the nature of the mediator $R$ and the allowed values DM parameters. In fact, it turns out to have only marginal impact on a scalar mediator but almost prohibits a vector mediation of missing-energy excess events by confining the vector-mediator mass $M_V$ to rather small values (as becomes clear from Fig. 4):

$$M_V \lesssim 3 \text{ GeV} \ . \tag{17}$$

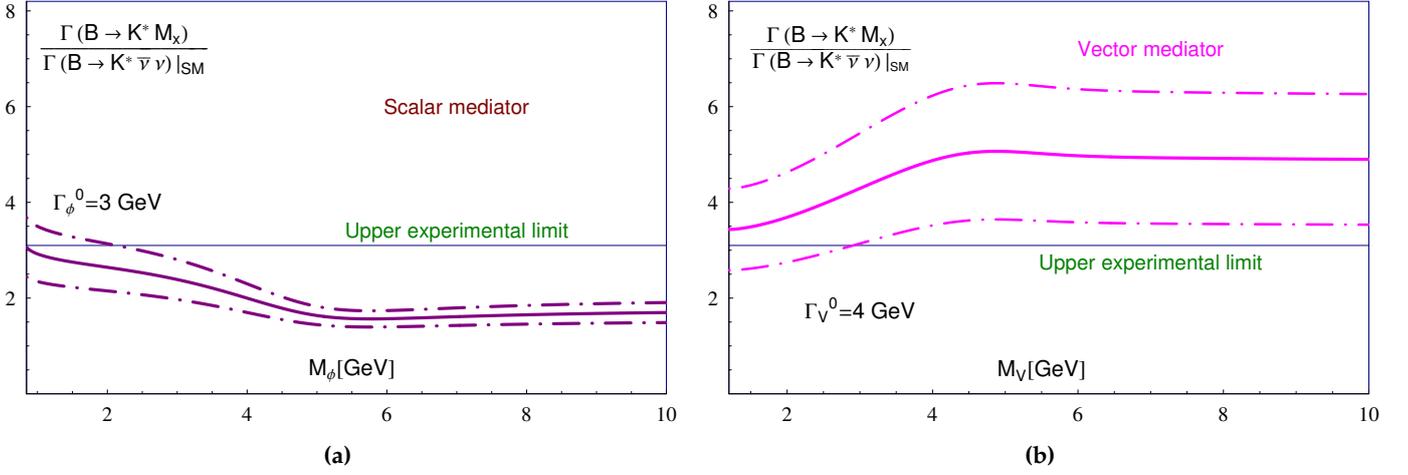

**Figure 4.** Expected relationship (15) (and errors) between integrated-width ratio (13) and ratio (12) [32] vs. mediator mass $M_{\phi,V}$, and related experimental upper limit (16), for (**a**) scalar or (**b**) vector mediator.

## 5. Interpreting Belle-II Missing-Energy Measurements $B$ Decay Findings

A minor problem encountered upon attempting to apply the considerations of Sect. 4 to the experimental data presented by the Belle II collaboration arises from the unfortunate fact that the Belle II experiment cannot reconstruct the direction of the $B$ mesons and, therefore, instead of the momentum difference $q$ squared has to use the "reconstruction" variable $q_{\text{rec}}^2$, in terms of the $B$-meson's energy $E_B$ and the $K$-meson's mass $M_K$ and energy $E_K$ defined by

$$q_{\text{rec}}^2 = E_B^2 + M_K^2 - 2\, E_B\, E_K \ . \tag{18}$$

The actual route along which all quantities defined and discussed in Sect. 4 as functions of $q^2$ are recalculated to functions depending on the variable $q_{\text{rec}}^2$ of Eq. (18), as required for their confrontation with the Belle-II outcomes, is described, in great detail, in Sect. III of Ref. [25].

The $\chi^2$ distributions of the *differential* decay rate $d\Gamma(B \xrightarrow{R} K\bar{\nu}\nu)/dq^2$ (extracted [25] from the information collected by the Belle II collaboration [1]) are depicted — both in the shape of two-dimensional landscapes as well as in the shape of the one-dimensional dependencies on essential DM parameters — for a scalar-mediator scenario in Fig. 5 and for a vector-mediator scenario in Fig. 6, where the location of the global minimum of $\chi^2$ is revealed by *the* black dot. Inferred from the position of the *global* minimum of $\chi^2$, the numerical values of the mass $M_R$ and decay width $\Gamma_R^0$ of the mediator state $R = \phi, V$ and the mass $m_\chi$ of the DM fermion $\chi$ are

$$M_\phi = 2.4 \pm 0.4 \text{ GeV}\, , \qquad \Gamma_\phi^0 = 2.9^{+1.1}_{-0.9} \text{ GeV}\, , \qquad m_\chi = 0.42^{+0.2}_{-0.4} \text{ GeV}\, , \tag{19}$$

$$M_V = 3 \text{ GeV}\, , \qquad \Gamma_V^0 = 4.0^{+2.0}_{-1.5} \text{ GeV}\, , \qquad m_\chi = 0.6^{+0.10}_{-0.18} \text{ GeV}\, , \tag{20}$$

where the *value* of $M_V$ is enforced by the upper bound (17); approximate numerical values of the mediator–dark-fermion couplings $g_{R\chi\chi}$ and the SM-quark–mediator couplings $g_{bsR}$ read

$$g_{\phi\chi\chi} \approx 6, \qquad g_{bs\phi} \approx 5 \times 10^{-8} \qquad \text{(scalar-mediator model of Subsect. 3.1)},$$
$$g_{V\chi\chi} \approx 7, \qquad g_{bsV} \approx 2 \times 10^{-8} \qquad \text{(vector-mediator model of Subsect. 3.2)}.$$

The comparison of the predictions emerging from our theoretical approach with the outcomes of the Belle II experiment [1] is, in at least two crucial respects, highly satisfactory:

1. The $q^2_{\text{rec}}$ distribution published by Belle II for the missing-energy excess events in the $B$ decay $B^+ \to K^+ M_X$ [1], as betrayed by both of the red curves in Fig. 7, might be easily reproduced both by the *scalar-mediator* model of Subsect. 3.1 and by the *vector-mediator* model of Subsect. 3.2 to a degree better than just acceptable. (This is not necessarily the case for less realistic choices for the numerical values of the involved DM parameters).

2. From the experimental point of view, it may be considered advisable to resort, for one's investigations, to the ("efficient") differential decay widths $\overline{d\Gamma^{\text{eff}}}(B \xrightarrow{R} K^{(*)} \bar\chi\chi)/dq^2_{\text{rec}}$, defined by weighting by the Belle-II detection efficiencies. However, the resulting ratio

$$\overline{\mathcal{R}^{(R)}_{K^*/K}}(q^2_{\text{rec}}) \equiv \frac{\dfrac{\overline{d\Gamma^{\text{eff}}}(B \xrightarrow{R} K^* \bar\chi\chi)}{dq^2_{\text{rec}}}}{\dfrac{\overline{d\Gamma^{\text{eff}}}(B \xrightarrow{R} K \bar\chi\chi)}{dq^2_{\text{rec}}}} , \qquad (21)$$

nevertheless, turns out to be nearly independent of any DM parameters and similar in its $q^2_{\text{rec}}$ behaviour (shown in Fig. 8) to the $q^2_{\text{rec}}$ dependence of the unweighted ratio (11):

$$\overline{\mathcal{R}^{(R)}_{K^*/K}}(q^2_{\text{rec}}) \approx \widehat{\mathcal{R}}^{(R)}_{K^*/K}(q^2_{\text{rec}}) .$$

## 6. Summary, Conclusions and Outlook: Selecting (Arguable) DM Models

For the two representatives of top-philic mediator-based DM models coarsely sketched in Sect. 3, the outcomes of the present comparative analysis may be summarized as follows:

1. The differential decay-width ratio (11), if considered as function of the missing energy, exhibits a continuous decrease to zero for the scalar-mediator scenario, and a moderate increase followed by an almost abrupt decline to zero for the vector-mediator scenario.
2. The integrated decay-width ratio (12) is, independently of the mediator mass, less than 1 for the scalar-mediator scenario, and larger than 1 for the vector-mediator scenario.
3. For the vector-mediator scenario only, the experimental upper limit on the ratio (13) of DM-over-SM missing-energy decays very tightly constrains the mediator-boson mass.
4. The experimental limitations or "boundary conditions" of the Belle-II measurements turn out not to have any significant impact on the trustability of our above conclusions.

These insights may constitute useful tools for identifying the nature of DM mediator bosons. Experimentally, the above predictions should be straightforwardly applicable to the existing or future findings of measurements by, e.g., Belle II. Theoretically, the applicability of the concepts proposed in Sect. 4 should be easily extendable to further categories of DM models.

If, for the decay $B \to K M_X$, the excess of missing-energy events can be corroborated and, for the decay $B \to K^* M_X$, the behaviour of the (differential) decay widths required for the application of our criteria gets established by future measurements, analyses of the kind exemplified here might prove to constitute a means to narrow down the conceivable spectrum of potential mediator-based DM scenarios. Moreover, there are good reasons to believe [32] that $B$-meson missing-energy decays to $\pi$ or $\rho$ mesons, $B \to (\pi, \rho) M_X$, should occur with quantitatively similar amount of excess events with respect to the SM decays $B \to (\pi, \rho) \bar\nu\nu$ to $\bar\nu\nu$ pairs and that for these decays analogous conclusions can be drawn [32]. The line of argument establishing this result is rather simple and unspectacular [32]: Each of

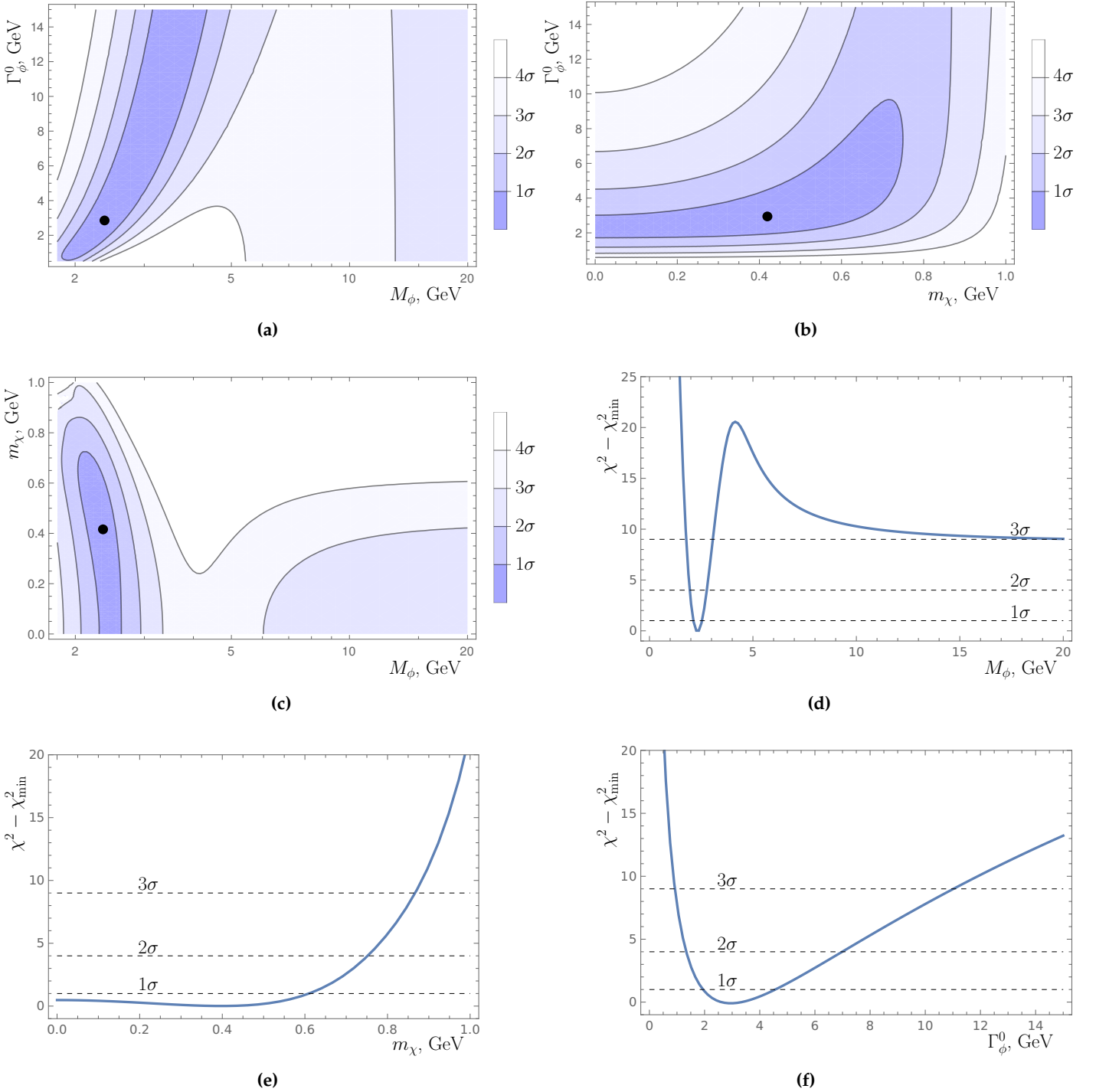

**Figure 5.** Two-dimensional $\chi^2$ landscapes for the pairings (**a**) $\Gamma_\phi^0$ vs. $M_\phi$, (**b**) $\Gamma_\phi^0$ vs. $m_\chi$, (**c**) $m_\chi$ vs. $M_\phi$ as well as dependencies of $\chi^2$ on (**d**) $M_\phi$, (**e**) $m_\chi$, (**f**) $\Gamma_\phi^0$, for the *scalar*-mediator scenario of Subsect. 3.1 [25].

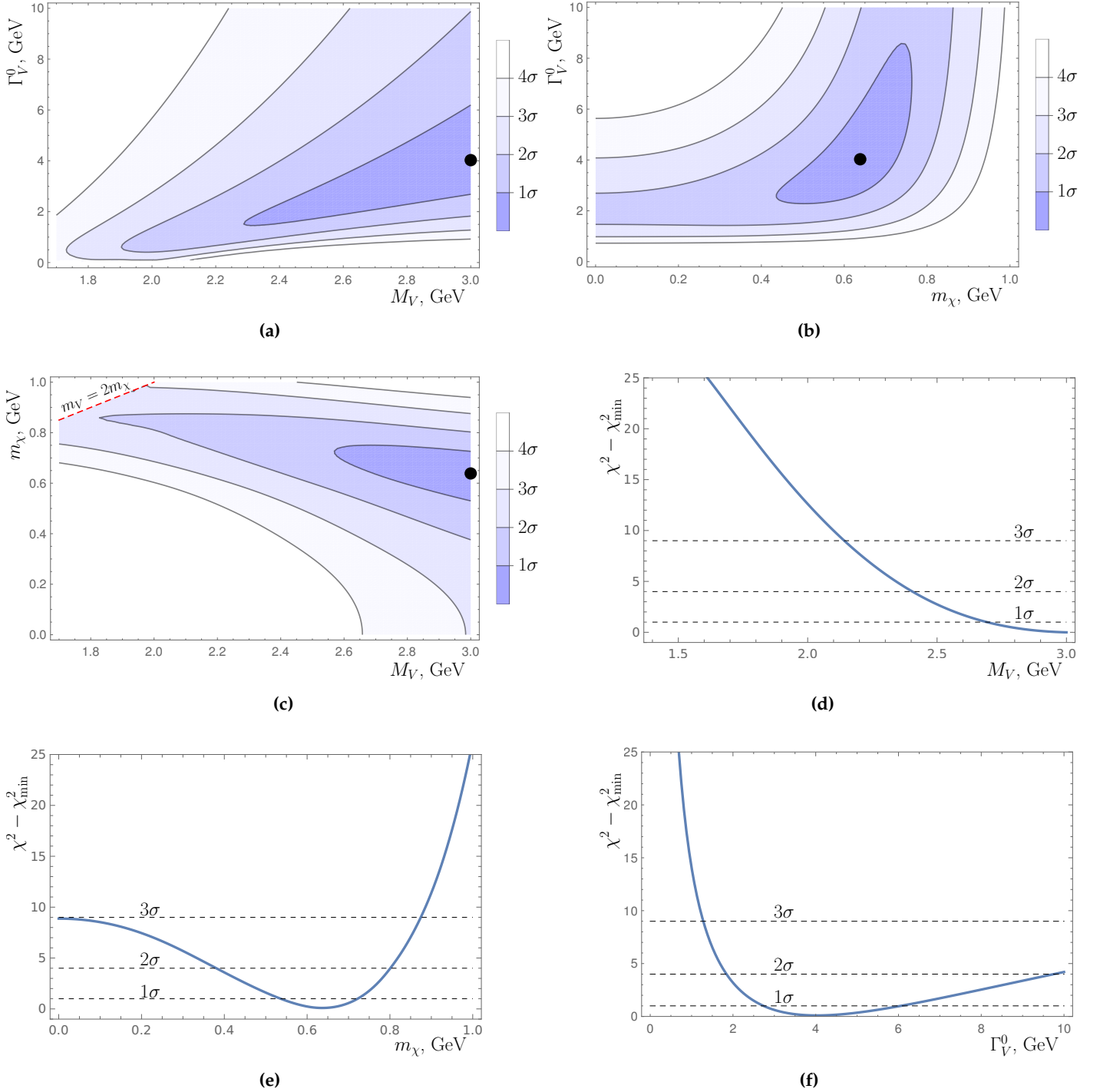

**Figure 6.** Two-dimensional landscapes of $\chi^2$ for its pairings (**a**) $\Gamma_V^0$ vs. $M_V$, (**b**) $\Gamma_V^0$ vs. $m_\chi$, (**c**) $m_\chi$ vs. $M_V$ as well as dependencies of $\chi^2$ on (**d**) $M_V$, (**e**) $m_\chi$, (**f**) $\Gamma_V^0$, for our *vector*-mediator scenario of Subsect. 3.2.

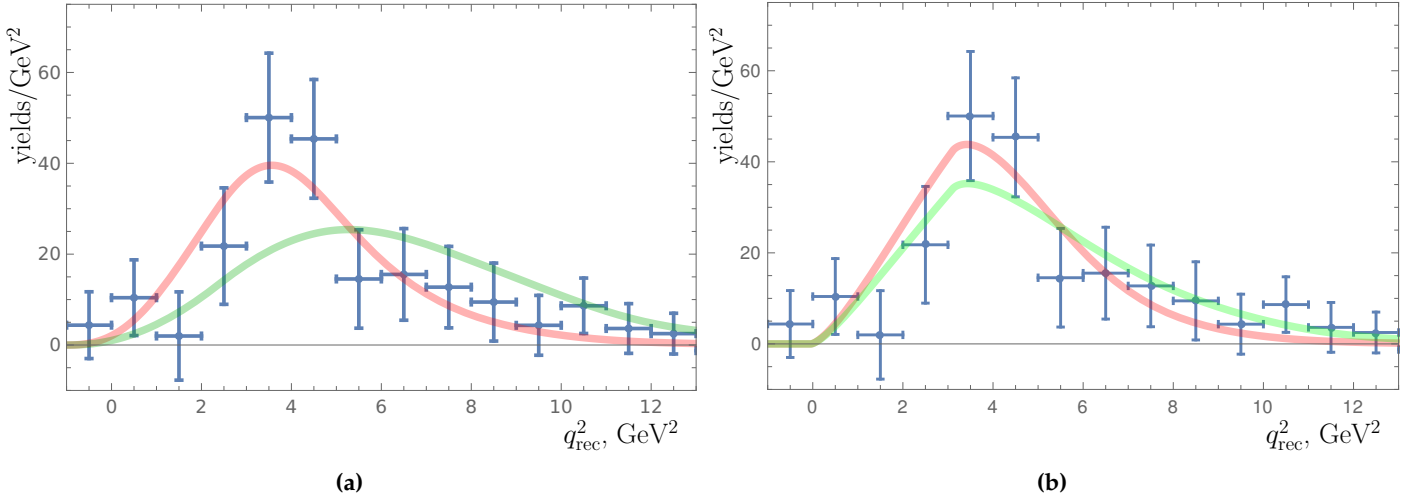

**Figure 7.** Reproduction of Belle-II results [1,13] for the decay $B \to K M_X$ by DM models with (**a**) a *scalar* mediator and either (red) the values (19) or (green) the not realistic values $M_\phi = 20$ GeV, $\Gamma_\phi^0 = 20$ GeV, $m_\chi = 0.42$ GeV of (basic) DM parameters [25] or (**b**) a *vector* mediator and either (red) the values (20) or (green) suboptimal values $M_V = 20$ GeV, $\Gamma_V^0 = 4$ GeV, $m_\chi = 0.6$ GeV of relevant DM parameters [32].

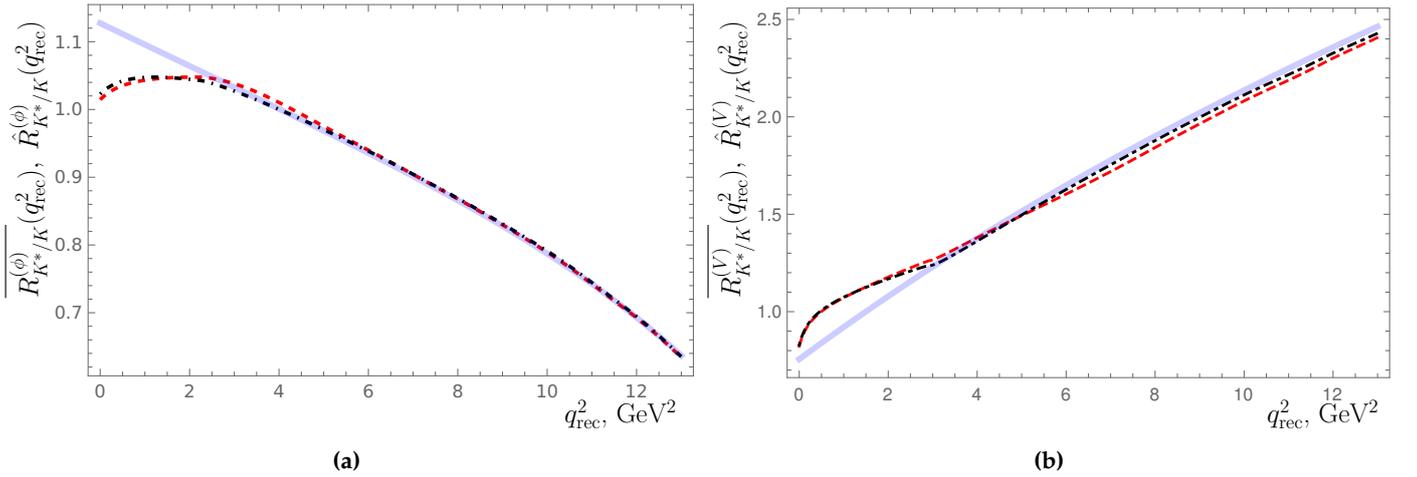

**Figure 8.** Theoretically predicted ratios (11) (blue) and experimentally *measurable* ratios (21) vs. $q_{\text{rec}}^2$, for (**a**) scalar [25] and (**b**) vector [32] mediators, emerging from either (red) the optimal values (19) and (20), respectively, or (black) fictitious large-$M_R$ sets of choices $M_\phi = 20$ GeV, $\Gamma_\phi^0 = 20$ GeV, $m_\chi = 0.42$ GeV for the *scalar*-mediator and $M_V = 20$ GeV, $\Gamma_V^0 = 20$ GeV, $m_\chi = 0.6$ GeV for the *vector*-mediator model.

12 of 14the decay widths under study involves the product of Cabibbo–Kobayashi–Maskawa matrix elements emerging from the two $W^\pm$-boson–quark vertices in Fig. 1. This product, however, clearly cancels in the ratio of (missing-energy over SM $\bar\nu\nu$) decay widths. Hence, apart from minor quark-mass effects the predicted decay-width ratios must be of similar or comparable magnitude for decays to *strange* and *nonstrange* (both either pseudoscalar or vector) mesons.

**Author Contributions:** Writing—original draft preparation, A.B., W.L. and D.M.; writing—review and editing, A.B., W.L. and D.M. All authors have read and agreed to the published version of the manuscript.

**Funding:** This research received no external funding.

**Institutional Review Board Statement:** Not applicable.

**Data Availability Statement:** Data sharing not applicable.

**Acknowledgments:** The research of A. B. and D. M. herein has been undertaken within the framework of the program "Particle Physics and Cosmology" of the National Center for Physics and Mathematics.

**Conflicts of Interest:** The authors declare no conflicts of interest.**Abbreviations**

The following abbreviations are used in this manuscript:

| | |
|---|---|
| BooNE | Booster Neutrino Experiment |
| FCNC | flavour-changing neutral current |
| HPQCD | High-Precision QCD |
| MDPI | Multidisciplinary Digital Publishing Institute |
| MILC | MIMD Lattice Computation |
| MIMD | Multiple Instruction, Multiple Data |
| PQCD | perturbative QCD |
| QCD | quantum chromodynamics |
| SM | Standard Model |
| SMEFT | SM effective field theory |
| Eq(s). | Equation(s) |
| Fig(s). | Figure(s) |
| Ref(s). | Reference(s) |
| Sect. | Section |
| Subsect(s). | Subsection(s) |

**Appendix A Simple Parametrization of Mesonic-Amplitude Form Factors**

Trustable parametrizations read for the *scalar-case* form factors $f_+^{B\to K}(q^2)$ and $f_0^{B\to K}(q^2)$

$$f_+^{B\to K}(q^2) = \frac{0.335}{(1 - q^2/M_{RV}^2)\,[1 - 0.58\,q^2/M_{RV}^2 + 0.03\,(q^2/M_{RV}^2)^2]}\,,$$
$$f_0^{B\to K}(q^2) = \frac{0.335}{1 - 0.648\,q^2/M_{RV}^2 - 0.17\,(q^2/M_{RV}^2)^2}\,,$$
$$M_{RV} = M_{B_s}(1^-) = 5.415\ \text{GeV}\,,$$

and for the four *vector-case* form factors $V^{B\to K^*}(q^2)$, $A_0^{B\to K^*}(q^2)$, $A_1^{B\to K^*}(q^2)$, and $A_2^{B\to K^*}(q^2)$

$$V^{B\to K^*}(q^2) = \frac{0.36}{(1 - q^2/M_{RV}^2)\,[1 + 0.15\,q^2/M_{RV}^2 + 0.166\,(q^2/M_{RV}^2)^2]}\,,$$
$$M_{RV} = M_{B_s}(1^-) = 5.415\text{ GeV}\,,$$
$$A_0^{B\to K^*}(q^2) = \frac{0.37}{(1 - q^2/M_{RS}^2)\,[1 - 0.656\,q^2/M_{RS}^2 + 0.03\,(q^2/M_{RS}^2)^2]}\,,$$
$$M_{RS} = M_{B_s}(0^-) = 5.366\text{ GeV}\,,$$
$$A_1^{B\to K^*}(q^2) = \frac{0.25}{(1 - q^2/M_{RA}^2)\,[1 + 0.097\,q^2/M_{RA}^2 + 0.02\,(q^2/M_{RA}^2)^2]}\,,$$
$$M_{RA} = M_{B_s}(1^+) = 5.829\text{ GeV}\,,$$
$$A_2^{B\to K^*}(q^2) = \frac{0.23}{(1 - q^2/M_{RA}^2)\,[1 - 0.114\,q^2/M_{RA}^2 - 0.128\,(q^2/M_{RA}^2)^2]}\,.$$